\documentstyle[preprint,eqsecnum,aps]{revtex}
\begin{document}
\preprint{SNUTP 97-166}
\tightenlines
\draft

\title{Ultrarelativistic limits of boosted dilaton black holes}

\author{ Rong-Gen Cai}
\address{ Center for Theoretical Physics, Seoul National University, 
Seoul 151-742, Korea}

\author{Jeong-Young Ji and Kwang-Sup Soh}
\address{ Department of Physics Education, Seoul National University,
 Seoul 151-742, Korea}

\maketitle

\begin{abstract}
We investigate the ultrarelativistic limits of dilaton black holes, 
black $p$-branes (strings), multi-centered dilaton black hole solutions
and black $p$-brane (string) solutions when the boost velocity
approaches the speed of light. For dilaton black holes and black
$p$-branes (boost is along the transverse directions), the resulting
geometries are gravitational shock wave solutions generated by a single
particle and membrane. For the multi-centered dilaton black
hole solutions and black $p$-brane solutions (boost is along the
transverse directions), the limiting  geometries are shock wave solutions
generated by multiple particles and membranes. When  the boost is along
the membrane directions, for the black $p$-brane and multi-centered
black $p$-brane solution, the resulting geometries describe general 
plane-fronted waves propagating along the membranes. The effect of the
dilaton on the limit is considered.

\end{abstract}
\pacs{{\it PACS}: 04.20.Jb, 04.30.Db, 04.30.Nk \\
{\it Keywords:} Dilaton black holes; Boost; Gravitational shock waves}

\newpage 

\section{introduction}

In recent years, much attention has been focused on 
 the ultrarelativistic limits of diverse black hole spacetimes. 
 There exist some reasons responsible for the interest. For example,
 people wish to know the spacetime structure of a moving black hole and
 use it to investigate the emission of gravitational waves when two
 black holes collide \cite{eath}. The ultrarelativistic boost of 
black holes usually leads to  a class  of gravitational wave spacetimes, 
which are exact solutions of Einstein's field equations. These resulting
gravitational shock wave solutions have been used to  study  
 the scattering of particles and strings at the energy of Planckian
scale \cite{hooft1,vega,ver,jack}. At that energy scale, just as pointed out 
by t' Hooft \cite{hooft1}, the gravitational interaction dominates their 
collision processes, the picture of particles (strings) propagating in a
flat spacetime ceases to be valid. The gravitational field generated by
the particles (strings) must be taken into account. 
In addition, they may  play an important role 
in understanding  the back reaction of Hawking radiation \cite{hooft2},
because particles with high velocity moving in curved spacetimes also 
produce gravitational shock waves.

In 1971, Aichelburg and Sexl (AS) \cite{as}
obtained an exact solution
of Einstein equation describing a particle moving at the speed of light
by boosting the Schwarzschild metric in the limit of the boost
velocity $v\rightarrow c=1$, where $c$ is the speed of light. The AS
geometry is flat everywhere, except for the location of the particle, where
the geometry has a delta-type singularity. The AS geometry  has been
explained as a shock wave spacetime generated by the null particle by
Dray and 't Hooft \cite{dray}, who have rederived the AS solution
by using the coordinate shift method. The coordinate shift method can
also be used to produce the shock wave in curved spacetimes, such as 
black hole spacetimes \cite{dray,lousto,sfetsos}. The shock wave solutions 
have been investigated in the higher derivative gravitational theory
\cite{buch,cam}.

The AS boost method has recently been  used to study the
ultrarelativistic limit of the Reissner-Nordstr\"om \cite{lousto1,stein}, 
Kerr \cite{kerr} and Kerr-Newman 
\cite{lousto2} spacetimes, and even the topological defect spacetimes 
\cite{lousto3}  and the Schwarzschild-(anti-)de
Sitter spacetimes \cite{hotta}. There some authors have investigated the
propagation and scattering of quantum fields in these ultrarelativistic
limiting geometries. Note that the geometric structure of black
holes is changed significantly due to the appearance of the dilaton field
 and the low energy actions
of superstring and supergravity theories can be viewed as the
modification of the Hilbert-Einstein action in general relativity.
In this paper, we would
like to investigate the effects of the dilaton field on the resulting 
ultrarelativistic geometry by boosting the four dimensional and higher
dimensional dilaton black holes to the ultrarelativistic limit.

In order to investigate 
the scattering process of multiple particles \cite{hooft1} and the back
reaction of Hawking radiation on the black hole geometry, it is
necessary to study the shock wave solution generated by multiple particles.
However, except for the shock wave solutions
generated by topological defects, those  solutions mentioned above 
 are all produced  by a single null particle (the shock wave in the de
 Sitter space is generated by two null particles \cite{hotta}).
By boosting the multi-centered dilaton black holes in the four
dimensional and higher dimensional spacetimes, we find the shock wave
solutions generated by multiple particles. Note that extremal black strings
and black $p$-branes are precisely solutions of fundamental strings and
membranes themselves. We get the ultrarelativistic geometries of black 
string and $p$-branes by boosting those solutions along transverse 
directions and
membrane (string) directions. Also we boost the multi-centered black
string and $p$-brane solutions and obtain gravitational wave spacetimes 
generated by multiple strings and membranes, respectively.

The organization of this paper is as follows. In the next section we
consider  boosting  a single four dimensional and higher dimensional dilaton
black hole. The case of multi-centered dilaton black hole solutions will be 
discussed in Sec.III. A  single black string and $p$-brane, and 
multi-centered black string and $p$-brane solutions 
are considered in Sec.IV and V, respectively. We present our conclusion  
with a discussion in Sec.VI.

\section{boosting the dilaton black holes}

\subsection{Four dimensional dilaton black holes}

Consider the following action
\begin{equation}
\label{action1}
S=\frac{1}{16\pi}\int d^4x\sqrt{-g}[R-2(\nabla \phi)^2-e^{-2a\phi}F_2^2],
\end{equation}
where $F_2$ is the Maxwell field. The coupling constant $a$ governs the
interaction between the dilaton field and the Maxwell field. For $a=1$,
the action is the low energy approximation of superstring action. The 
charged dilaton black hole solutions are \cite{garf}
\begin{eqnarray}
&& F_{tr}=\frac{e^{2a\phi}}{R^2}Q, \\
&&  e^{2a\phi}=\left( 1-\frac{r_-}{r}\right)^{2a^2/(1+a^2)}, \\
\label{met1}
&& ds^2=-A^2(r)dt^2 +A^{-2}(r)dr^2 +R^2(r)d\Omega_2^2,
\end{eqnarray}
where
\begin{eqnarray}
&& A^2(r)=\left(1-\frac{r_+}{r}\right)\left
       (1-\frac{r_-}{r}\right)^{(1-a^2)/(1+a^2)},\\
&& R^2(r)=r^2\left(1-\frac{r_-}{r}\right)^{2a^2/(1+a^2)}.
\end{eqnarray}
The mass and electric charge of the hole are
\begin{equation}
 2M=r_+ +\frac{1-a^2}{1+a^2}r_-, \ \ \ \ \  Q^2 =\frac{r_- r_+}{1+a^2}.
\end{equation}
For our purposes, we rewrite the metric (\ref{met1}) in the isotropic
coordinates. Defining 
\begin{equation}
r=\bar{r}\left( 1+\frac{r_+
   +r_-}{2\bar{r}}+\frac{(r_+-r_-)^2}{16\bar{r}^2}\right),
\end{equation}
we have
\begin{equation}
ds^2=-A^2(r)dt^2 +\bar{r}^{-2}R^2(r)[d\bar{r}^2 +\bar{r}^2 d\Omega_2^2].
\end{equation}
Here $\bar{r}$ is asymptotically a Cartesian coordinate and one 
has $\bar{r}^2= x^2 +y^2 +z^2$. Performing a Lorentz transformation
in the $x$-direction:
\begin{eqnarray}
\label{lorentz}
&& t=\gamma (t'-v x'),\ \  \ \ y=y', \nonumber \\
&& x=\gamma (x'-v t'),\ \  \ \ z=z', 
\end{eqnarray}
where $\gamma =(1-v^2)^{-1/2}$ and the constant $v$ is the boost
velocity, one has  
\begin{equation}
\label{met2}
ds^2=\gamma ^2 [(\bar{r}')^{-2}R^2(\bar{r}')-A^2(\bar{r}')](dt'-vdx')^2
      +(\bar{r}')^{-2} R^2(\bar{r}')(-dt'^2 +dx'^2 +dy'^2 +dz'^2),
\end{equation}
where $\bar{r}'^2=\gamma ^2 (x'-vt')^2 +y'^2 +z'^2$. We  expand the
components of the
metric (\ref{met2}) up to the order of $r_+^2$, $r_-^2$, and $r_-r_+$
(higher-order contributions will vanish due to the boost). The
results are
\begin{eqnarray}
A^2(\bar{r}') &=& 1-\left( r_+ +\frac{1-a^2}{1+a^2}r_-
        \right)\frac{1}{\bar{r}'}    \nonumber \\
			 && +\left[ \frac{r_+(r_++r_-)}{2}
			 +\frac{1-a^2}{1+a^2}\frac{r_-(r_++r_-)}{2}
			 -\frac{a^2(1-a^2)}{(1+a^2)^2}r_-^2 
			 +\frac{1-a^2}{1+a^2}
			 r_+r_-\right]\frac{1}{\bar{r}'^2}, \\
(\bar{r}')^{-2} R^2  &=& 1+\left( r_+ +\frac{1-a^2}{1+a^2}r_-
            \right )\frac{1}{\bar{r}'}   \nonumber \\
            && +\left [\frac{3r_+^2 +3r_-^2 +2
				r_+r_-}{8}-\frac{a^2r_+r_-}{1+a^2} -\frac{2a^2
				r_-^2}{(1+a^2)^2}\right ]\frac{1}{\bar{r}'^2}.
\end{eqnarray}				
We now take the limit $v\rightarrow 1.$  This limit is a delicate point.
Employing the method in Ref.\cite{as} (see also Ref.\cite{lousto1}), we can 
obtain the results of physical meaning. This method works as follows.
One integrates first the expression with respect to $u'=x'-vt'$, 
takes the limit $v\rightarrow 1$, and then differentiates  the corresponding
expression with respect to $u'$. In order that the black hole
can be boosted to the velocity of light, the mass and charge  must go
to zero. Following \cite{as,lousto1,lousto2}, we set
\begin{equation}
\label{mrelation}
M=\gamma ^{-1}p, \ \ Q^2=\gamma ^{-1}p_e^2,
\end{equation}
where $p$ and $p_e$ are two constants and can be interpreted  as the
 kinematic  and the electromagnetic momenta, respectively. 
Thus we can obtain
\begin{eqnarray}
ds^2 &=&\left \{-4p\left [\delta (x'-t')\ln \rho ^2-\frac{1}{|t'-x'|} 
    \right ] -\frac{3-4a^2}{2(1-a^2)} \frac{\pi
	 p^2_e}{\rho}\delta (x'-t')\right\}(dt'-dx')^2 \nonumber \\
	&+& (-dt'^2 +dx'^2 +dy'^2 +dz'^2), 
\end{eqnarray}	 
where $\rho ^2=y'^2 +z'^2$. Performing the following coordinate
transformation 
\begin{eqnarray}
&& dy'=dy, \ \ dz'=dz, \nonumber \\
&& dt'-dx' =dt- dx, \nonumber \\
&& dt'+dx'= dt + dx -\frac{4p}{|t'-x'|}(dt-dx),
\end{eqnarray}
we have 
\begin{equation}
\label{solution1}
ds^2 = dudv +dy^2 +dz^2                                     
     -\left\{ 4p \ln \rho ^2 +
  \frac{3-4a^2}{2(1-a^2)} 
	  \frac{\pi p^2_e}{\rho}\right\}\delta (u) du^2,
\end{equation}
where   
$ u=x-t, \ \ v=x+t $
are two null coordinates.
When $a=0$, the solution (\ref{solution1}) reduces to the one 
for the Reissner-Nordstr\"om black holes \cite{lousto1}:
\begin{equation}
\label{rn}
ds^2=dudv +dy^2 +dz^2 -\left[4p\ln \rho^2 +\frac{3}{2}\frac{\pi
p^2_e}{\rho}\right]\delta (u)du^2.
\end{equation}
From the result (\ref{solution1}) it is easy to find
that the contribution of the dilaton field enters the expression
through the contribution of the electromagnetic field. This is expected
because the dilaton field has to be zero when the electromagnetic field 
disappears. In that case the solutions (\ref{solution1}) and (\ref{rn}) 
both reduce to 
\begin{equation}
\label{assolu}
ds^2=dudv +dy^2 +dz^2 -4p\ln \rho^2 \ \delta (u) du^2.
\end{equation}
This is the AS solution obtained by boosting the Schwarzschild black 
hole.  From (\ref{solution1}) we find that when $a^2=3/4$, the
contributions of the dilaton and electric fields are canceled each other,
and the resulting limit is the AS solution (\ref{assolu}). This is a
very interesting case. In addition, it is worth noting that if we set 
$|Q|$ go to zero as the mass, that is 
\begin{equation}
\label{mqrelation}
Q^2=\gamma ^{-2}p^2_e,
\end{equation}
both  the contributions of the electric field and dilaton field vanish,
 and the resulting metric is
just the AS solution.

In eq.~(\ref{mrelation}) the first relation has a 
good motivation \cite{as}; the second relation is to have a
distributionally well defined limit of the metric.  But, in  this case,
the electromagnetic field has a vanishing field strength (the gauge
potenial is a pure gauge \cite{jack}) but a nonzero,
delta-like energy density in the ultrarelativistic limit 
\cite{lousto1}.  It is easy to prove that the dilaton field has also 
the similar properties.   However, this situation is still 
mathematically perfectly  defined \cite{stein}.

 From (\ref{solution1}), it is obvious that 
the solution is invalid as $a=1$. The case $a=1$ is a special one. In
this case, we have $r_+=2M$, $r_-=Q^2/M$, and 
\begin{eqnarray}
A^2 &=& 1-\frac{r_+}{\bar{r}'}+\frac{r_+(r_++r_-)}{2\bar{r}'^2}, \\
(\bar{r}')^{-2} R^2 &=& 1+\frac{r_+}{\bar{r}'}+ \left[ \frac{(r_++r_-)^2}
    {4} -\frac{r_-(r_++r_-)}{2}+\frac{(r_+-r_-)^2}{8}\right]\frac{1}
		 {\bar{r}'^2}.
\end{eqnarray}
If one still uses the relations in eq.~(\ref{mrelation}), he will find
that the contribution of the dilaton  field is divergent. 
To gain a well defined limit of this metric, we have to rescale the mass
and charge as
\begin{equation}
\label{qrelation}
M=\gamma^{-1}p, \ \ Q^2=\gamma ^{-3/2}p^2_e.
\end{equation}
Using the same method, we get the ultrarelativistic limit 
\begin{equation}
\label{asolu}
ds^2=dudv +dy^2 +dz^2 -\left [ 4p \ln \rho ^2 +\frac{p^2_e}{8p^2}
   \frac{\pi p^2_e}{\rho}\right ] \delta (u) du^2.
\end{equation}
This situation happens due to the fact that  
the energy-momentum tensor of the dilaton field is proportional to
$r^2_-=Q^4/M^2$, but the one of the electric field is proportional 
to $Q^2$. The second term in the square brackets in (\ref{asolu}) comes
from the contribution of the dilaton field and the electric field has
not any contribution in this case.

\subsection{Higher dimensional dilaton black holes}

Now we extend the above discussion to the arbitrary higher dimensional
dilaton black holes, which come from the action \cite{horo}
\begin{equation}
S=\frac{1}{16\pi}\int d^Dx\sqrt{-g}\left[R-\frac{1}{2}(\nabla \phi)^2 
      -\frac{2e^{\beta \phi}F_{D-2}^2}{(D-2)!}\right],
\end{equation}
where $F_{D-2}$ is a $(D-2)$-form satisfying $dF=0$ and $\beta$ is a
coupling constant. The black hole solutions are
\begin{eqnarray}
 F_{d+1} &=& Q\epsilon _{d+1}, \\
 e^{\beta \phi} &=& \left [1-(r_-/r)^d\right ]^{d b}, \\
 \label{met3}
 ds^2 &=&- A (r) dt^2 +B (r)dr^2 +r ^2 C(r)d\Omega _{d+1},
 \end{eqnarray}
 where 
 \begin{eqnarray}
 && A(r)=\left[1-(r_+/r)^d\right] \left [1-(r_-/r)^d
           \right]^{1-db}, \nonumber \\
 && B(r)= \left[1-(r_+/r)^d\right]^{-1}\left[ 1-
       (r_-/r)^d\right] ^{b-1}, \nonumber \\
&& C(r)=\left[1-(r_-/r)^d\right]^b, \nonumber 
\end{eqnarray}		
 $d=D-3$ and  the constant $b$ is
$$ b=\frac{2(d+1)\beta ^2}{d [2d +\beta ^2 (d+1)] }.$$
The magnetic charge $Q$ satisfies 
\begin{equation}
Q^2 =\frac{b d^3  (r_+r_-)^d}{2 \beta ^2}.
\end{equation}
 The mass of the black hole is 
 \begin{equation}
 M=\frac{(d+1)A_{d+1}}{16\pi}[r_+^d +(1-d b)r_-^d].
 \end{equation}
Here $A_{d+1}=2\pi ^{(d+2)/2}/\Gamma ((d+2)/2)$ is the volume of the 
unit $(d+1)$-sphere. Rewriting
the metric (\ref{met3}) in the isotropic coordinates yields
\begin{equation}
\label{met4}
ds^2 =-A(r)dt^2 +\bar{r}^{-2}r^2 C(r)[d\bar{r}^2 +\bar{r}^2d\Omega
         ^2_{d+1}]
\end{equation}			
where 
\begin{equation}
\label{relation}
r^d =\bar{r}^d \left[ 1+\frac{r_+^d+r_-^d}{2\bar{r}^d }  
     +\frac{(r_+^d-r_-^d)^2}{16\bar{r}^{2d}}\right ].
\end{equation}
Expanding the functions $A$ and $r^2(\bar{r})^{-2}C$, up to the order of
$r_+^{2d}$, $r_-^{2d}$ and $r_+^dr_-^d$, we obtain
\begin{eqnarray}
A &=& 1-\left[r_+^d +(1-bd)r_-^d\right]\frac{1}{\bar{r}^d}
    +\left[\frac{r_+^d(r_+^d+r_-^d)}{2}
	 \right. \nonumber \\
	 &&\left. +\frac{bd(bd-1)r_-^{2d}}{2}
	  -\frac{(bd-1)r_-^d(r_+^d+r_-^d)}{2}-(bd-1)r_+^dr_-^d\right]\frac{1}
	  {\bar{r}^{2d}}, \\
r^2(\bar{r})^{-2}C &=& 1+\left[\frac{r_+^d+r_-^d}{d}-br_-^d\right]
         \frac{1}{\bar{r}^{d}} 
			+\left[
			\frac{(r_+^d-r_-^d)^2}{8d}-\frac{br_-^d(r_+^d+r_-^d)}{d}
		   \right. \nonumber \\
			&& \left.+\frac{(2-d)(r_+^d+r_-^d)^2}{4d^2}
			+\frac{br_-^d(r_+^d+r_-^d)}{2}
			+\frac{b(b-1)r_-^{2d}}{2}\right]\frac{1}{\bar{r}^{2d}},
\end{eqnarray}
where $\bar{r}^2=(x^1)^2+(x^2)^2 +\cdots +(x^{d+2})^2$. 
Now boosting the dilaton black hole in a direction, say $x^1$, and
taking the limit $v\rightarrow 1$, we obtain
\begin{eqnarray}
\label{hd}
ds^2 &=& dudv + (dx^2)^2 +(dx^3)^2 +\cdots +(dx^{d+2})^2 \nonumber \\
     &+& \left [\frac{16\pi p}{(d-1)A_d \ \rho ^{d-1}} + 
	  \frac{ f p^2_e}{\rho ^{2d-1}}\right]\delta (u) du^2,
\end{eqnarray}
where $ u=x^1-t$, $v=x^1+t$, the constant $f$ is
\begin{equation}
\label{fsolution}
f=\left [\frac{(db-2)(2-3db)}{8d(1-db)}+\frac{3db-2}{2}
  \right]\frac{2d+\beta ^2(d+1)}{d^2(d+1)}\frac{\pi
  (2d-3)!!}{2^{d-1}(d-1)!},
\end{equation}
and 
$$ \rho ^2 =(x^2)^2 +(x^3)^2 +\cdots +(x^{d+2})^2. $$
It should be pointed out that in the solution (\ref{hd}) $d>1$. When
$\beta =0$, i.e., $b=0$, the solution (\ref{hd}) gives the result of the
higher dimensional Reissner-Nordstr\"om black hole \cite{lousto1}. When 
the charge $Q$ vanishes, the solution (\ref{hd}) reduces to
\begin{equation}
\label{has}
ds^2=dudv +(dx^2)^2 +(dx^3)^2 +\cdots +(dx^{d+2})^2 +\left [ 
     \frac{16\pi p}{(d-1)A_d \ \rho^{d-1}}\right] 
	  \delta (u)du^2.
\end{equation}
This is the  higher dimensional extension of the AS solution. 
From (\ref{fsolution}) one can see that when $b=2/3d$, i.e., $\beta^2=
d/(d+1)$, the contributions of the dilaton and tensor fields are
canceled.  In this case the resulting limit is the solution 
(\ref{has}) again. In addition,
when $b=1/d$, i.e., $\beta^2=2d/(d+1)$, 
the metric (\ref{hd}) is invalid. Like the case $a=1$ in four
dimensions, we have to use the scaling relation
(\ref{qrelation}) in this case, and then the resulting  metric  
is the line element (\ref{hd}) with 
\begin{equation}
f=-\frac{2p^2_e A^2_{d+1}}{(16\pi)^2 d^3 p^2}\frac{\pi (2d-3)!!}
  {2^{d-1}(d-1)!}.
\end{equation}

\section{Boosting the multi-centered dilaton black hole solutions}

\subsection{Four dimensional multi-centered dilaton 
black hole solutions}

In the action (\ref{action1}),
 when the extremality or zero-force condition is satisfied,
besides the extremal dilaton black hole solution (\ref{met1}) with 
$r_+=r_-$, one has the so-called multi-centered dilaton black 
hole  solutions\cite{shi},
\begin{equation}
ds^2=-U^{-2}(r)dt^2 +U^2(r) (dx^2 +dy^2 +dz^2),
\end{equation}
where
\begin{equation}
U(r)=\left (1+\sum^{n}_{i=1}\frac{\mu_i}{|r-r_i|}\right )^{1/(1+a^2)},
\end{equation}
$r_i=(x_i,y_i,z_i)$ is the location of the $i$th black hole. 
The potential and the dilaton configuration are
\begin{eqnarray}
&& A=\pm \left(\frac{1}{1+a^2}\right)^{1/2}
    \left (1+\sum^{n}_{i=1}\frac{\mu_i}{|r-r_i|}\right )^{-1}dt,\\
&& e^{-2a\phi}=	 
    \left (1+\sum^{n}_{i=1}\frac{\mu_i}{|r-r_i|}\right )^{2a^2/(1+a^2)}.
\end{eqnarray}
 This solution describes $n$ extremal dilaton black holes in 
static equilibrium. The constant $\mu_i$ has the relation to the mass 
of $i$th  black hole as $\mu _i=(1+a^2)m_i $. When $a=0$, this solution 
reduces to the Majumdar-Papapetrou (MP) solution \cite{mp}, which 
describes $n$
extremal Reissner-Nordstr\"om black holes in the equilibrium. 

Now we boost the multi-centered dilaton black holes to the
ultrarelativistic limit. Using the Lorentz boost (\ref{lorentz}),
 we have 
\begin{equation}
\label{metric}
ds^2=\gamma ^2 [U^{2}(r')-U^{-2}(r')](dt'-vdx')^2 +
     U^{-2}(r')(-dt'^2 +dx'^2 +dy'^2 +dz'^2),
\end{equation}
where $r'^2= \gamma^2 (x'-vt')^2 +y'^2 +z'^2$.
Taking the limit $v\rightarrow 1$, in order to get a solution of
physical meaning, one has to make $m_i$ change as \cite{as}
\begin{equation}
\label{mmrelation}
m_i=p_i\gamma ^{-1},
\end{equation}
where $p_i$  remains 
unchanged in the process of the boost. Expanding the components of 
metric (\ref{metric}) up to the order $m_i$, yields
\begin{eqnarray}
&& U^2(r)=1 +\frac{2}{1+a^2}\sum^{n}_{i=1}\frac{\mu_i}{|r-r_i|},\\
&& U^{-2}(r)=1- \frac{2}{1+a^2}\sum^{n}_{i=1}\frac{\mu_i}{|r-r_i|},
\end{eqnarray}
Taking the limit $v\rightarrow 1$, we obtain
\begin{equation}
ds^2=dudv +dy^2 +dz^2 -\sum^{n}_{i=1}4p_i \left (\delta (u-u_i)\ln 
\rho ^2_i+ \frac{1}{|u-u_i|}\right )du^2,
\end{equation}
where we have dropped the prime, and
\begin{eqnarray}
&& u=x-t, \ \ v=x+t, \\
&& \rho ^2_i= (y-y_i)^2 +(z-z_i)^2.
\end{eqnarray}
Performing the following coordinate transformation 
\begin{equation}
du \rightarrow du, \ \ dv \rightarrow dv
-\sum^{n}_{i=1}\frac{4p_i}{|u-u_i|}du, 
\end{equation}
yields the result we expect
\begin{equation}
\label{mas}
ds^2=dudv +dy^2 +dz^2 -\sum^{n}_{i=1}4p_i\ln \rho^2_i 
\ \delta (u-u_i) du^2.
\end{equation}
This solution is independent of the coupling parameter
$a$. So the ultrarelavisitic limit of the MP solution is also the
equation (\ref{mas}). From  the solution, the nonvanishing component of 
the stress-energy tensor is
\begin{equation}
T_{uu}=\sum^{n}_{i=1}p_i\delta (u-u_i)\delta (y-y_i)\delta (z-z_i),
\end{equation}
this is the sum of tensors of the $n$ independently null particles.
Therefore  the
shock wave spacetime  (\ref{mas}) is generated by $n$  particles
moving along a same direction  at the speed of light. No
interaction between $n$ particles is related to the fact that the
multi-centered  
solution describes the spacetime where $n$ extremal 
black holes are in the static equilibrium. Therefore it 
 is an extension of AS geometry to the
multiple particles. When $n=1$ and $x_1=y_1=z_1=0$, obviously, the solution
(\ref{mas}) goes back  to the AS solution. Similar to the AS geometry, our
solution (\ref{mas}) is flat except for those surfaces where particles live.

\subsection{Higher dimensional multi-centered dilaton black hole
solutions}

We now generalize the above solution to the higher dimensional
spacetimes. Consider the action \cite{shi}
\begin{equation}
\label{action3}
S=\frac{1}{16\pi}\int d^Dx\sqrt{-g}\left [R-\frac{4}{D-2}
    (\nabla \phi)^2 -e^{-4a\phi /(D-2)}F^2_2\right ],
\end{equation}
where $F_2$ is also the Maxwell field. In this case, the multi-centered
solution is
\begin{eqnarray}
\label{hmp}
&& ds^2 =-U^{-2}(r)dt^2 +U^{2/d}(r)\delta _{ab}dx^adx^b, \\
&& A=\pm \left (\frac{d+1}{2(d+a^2)}\right )^{1/2} F^{-1}(r)dt, \\
&& e^{-4a\phi /(D-2)}=[F(r)]^{2a^2/(d+a^2)}.
\end{eqnarray}
The function $U(r)$ is
\begin{equation}
U(r)=[F(r)]^{d/(d+a^2)},
\end{equation}
with
$$ F(r)=1+\frac{1}{d}\sum^{n}_{i=1}\frac{\mu_i}{|r-r_i|^d}.$$
Here $d=D-3$, $a,b=1,2, \cdots, d+2$, and 
$$ r^2=(x^1)^2 +(x^2)^2 +\cdots +(x^{d+2})^2, \ \ \ 
 r_i^2=(x^1_i)^2 +(x^2_i)^2 +\cdots +(x^{d+2}_i)^2.$$
 The constant $\mu_i$ is related to the mass of the $i$th hole as
\begin{equation}
\mu_i=\frac{8\pi (d+a^2)}{(d+1)A_{d+1}}m_i.
\end{equation}
When $a=0$, the solution (\ref{hmp}) is just the higher dimensional
generalization of the MP solution given by Myers \cite{myers}. Now we
boost the multi-centered solution along the $x^1$ direction and use the 
relation (\ref{mmrelation}). The resulting metric is 
\begin{equation}
ds^2 =dudv +(dx^2)^2 +(dx^3)^2 +\cdots +(dx^{d+2})^2 
      +\sum^{n}_{i=1}\frac{16\pi p_i}{(d-1)A_d \ \rho _i^{d-1}}\delta
		(u-u_i)du^2,
\end{equation}
with
\begin{equation}
\label{rho_i}
 \rho^2 _i=(x^2-x_i^2)^2 +(x^3 -x_i^3)^2+\cdots +(x^{d+2}-x_i^{d+2})^2.
\end{equation}
Again, the solution is independent of the coupling constant $a$. 
Therefore this is also the ultrarelavisitic limit of the Myers' 
solution. This metric is the higher dimensional generalization of 
the solution (\ref{mas}).

\section{Boosting the dilaton black strings and $p$-branes}

Now let us consider the low energy action of supergravity theory
\begin{equation}
\label{action4}
S=\frac{1}{16\pi}\int d^Dx \sqrt{-g}\left [ R-\frac{1}{2}(\nabla \phi)^2
  -\frac{e^{-\alpha (d)\phi}}{2(d+1)!}F^2_{d+1}\right],
\end{equation}
where $F_{d+1}$ denotes the $(d+1)$-form tensor field, and 
$$ \alpha ^2(d)=4-\frac{2d\tilde{d}}{d +\tilde {d}}, \ \ \tilde
{d}=D-d-2.$$
In the action (\ref{action4}) one has the black $(p=\tilde{d}-1)$-brane
(black string for $p=1$) solution with ``magnetic'' charge \cite{duff}
\begin{eqnarray}
&& F=Q\epsilon _{d+1}, \\
&& e^{-2\phi}=\left[1-(r_-/r)^d\right]^{\gamma_{\phi}}, \\
\label{bp}
&& ds^2=-A(r)dt^2 +B(r)dr^2 +r^2C(r)d\Omega _{d+1}+ D(r)
    \delta _{ab}dy^a dy^b,
\end{eqnarray}
where
\begin{eqnarray}
&& A(r)=\left[1-(r_+/r)^d \right]\left[1-(r_-/r)^d\right
          ]^{\gamma_{y}-1},   \nonumber \\
&& B(r)=\left[1-(r_+/r)^d\right]^{-1}\left [1-(r_-/r)^d\right]^
     {\gamma _{\Omega}-1},  \nonumber \\
&& C(r)=\left[1-(r_-/r)^d\right ]^{\gamma _{\Omega}}, \nonumber \\
&& D(r)=\left [1-(r_-/r)^d\right ]^{\gamma _y}, \nonumber 
\end{eqnarray}
and 
$$ \gamma _y=\frac{d}{d+\tilde d}, \ \ \gamma
_{\Omega}=\frac{\alpha^2(d)}{2d}, \  \ \gamma_{\phi}=\alpha (d).$$
The black $(d-1)$-brane solution with ``electric'' charge in the action 
(\ref{action4}) can be obtained by using the duality transformation
\cite{cai}. In
the solution (\ref{bp}) the charge $Q$ and the ADM mass $M$ per unit
$p$-brane are
\begin{eqnarray}
&& Q=d(r_-r_+)^{d/2}, \\
&& M=\frac{A_{d+1}}{16\pi}[(d+1)r^d_+-r^d_-].
\end{eqnarray}
There exist two different kinds of directions in the $p$-brane solutions
(\ref{bp}): transverse direction and membrane direction. We will boost
the solution along the two kinds of  directions, respectively.

\subsection{Boosting in the transverse  directions}

Rewriting the black $p$-brane (\ref{bp}) in the isotropic coordinates
yields 
\begin{equation}
\label{bp2}
ds^2=-A(r)dt^2 +r^2\bar{r}^{-2}C(r)[d\bar{r}^2 +\bar{r}^2
    \Omega^2_{d+1}] +D(r)\delta_{ab}dy^ady^b,
\end{equation}
where the relation between $r$ and $\bar{r}$ is the same as the one in
(\ref{relation}). Expanding the functions $A(r)$ and
$r^2\bar{r}^{-2}C(r)$ up to the order of $r^{2d}_+$, $r_-^{2d}$ and
$r_+^dr_-^d$, one has
\begin{eqnarray}
A(r) &=& 1-\left[r_+^d +(\gamma _y -1)r_-^d\right]\frac{1}{\bar{r}^d} 
			+\left[  \frac{r_+^d(r_+^d+r_-^d)}{2} 
			+\frac{(\gamma_y-1)r_-^d(r_+^d+r_-^d)}{2}
			\right. \nonumber \\
         &+& \left.\frac{(\gamma_y-1)(\gamma_y-2)r_-^{2d}}{2}
			+(\gamma_y-1)r_+^dr_-^d\right]\frac{1}{\bar{r}^{2d}},\\
r^2\bar{r}^{-2}C(r) &=&1+\left[\frac{r_+^d+r_-^d}{d}-
              \gamma_{\Omega}r_-^d \right]\frac{1}{\bar{r}^d}			
				 +\left [\frac{(r_+^d-r_-^d)^2}{8d}-\frac{\gamma_{\Omega}
				 r_-^d(r_+^d+r_-^d)}{d} 
				 \right. \nonumber \\
				 &+& \left. \frac{(2-d)(r_+^d+r_-^d)^2}{4d^2} 
				 +\frac{\gamma_{\Omega}r_-^d(r_+^d+r_-^d)}{2}
				 +\frac{\gamma_{\Omega}(\gamma_{\Omega }-1)r_-^{2d}}{2}
				 \right]\frac{1}{\bar{r}^{2d}}.
\end{eqnarray}
Using the relation (\ref{mrelation}) and taking the limit 
$v\rightarrow 1$, we obtain 
the resulting solution due to the boost along a direction, say $x^1$, of 
the transverse directions 
\begin{eqnarray}
\label{sbp}
ds^2 &=& dudv +(dx^2)^2 +(dx^3)^2 +\cdots +(dx^{d+2})^2
            +\delta_{ab}dy^ady^b \nonumber \\
		&+&\left [\frac{16\pi p}{(d-1)A_d\ \rho ^{d-1}} +
		\frac{g\  p_e^2}{\rho^{2d-1}}\right]\delta (u)du^2,
\end{eqnarray}
where $u=x^1-t$, $v=x^1+t$, and the constant $g$ is
\begin{eqnarray}
 g =&& \left[\frac{1}{8d(d+1)}+
         \frac{(d^2-4)\gamma_{\Omega}}
      {2d^3}-\frac{(d+2)(4-3d^2)}{4d^4(d+1)}\right. \nonumber \\
		&+& \left. \frac{\gamma_{\Omega}
	(\gamma_{\Omega}-1)(d+1)}{2d^2}-\frac{(\gamma_y-1)
   	[(d+1)\gamma_y+d+4] }{2d^2}\right]
		\frac{\pi (2d-3)!!}{2^{d-1}(d-1)!}. \nonumber
\end{eqnarray}		
The solution (\ref{sbp}) is the gravitational shock wave spacetime
generated by the membrane moving at the speed of light along the 
direction perpendicular to the  membrane itself. If we use the relation 
(\ref{mqrelation}) or let $Q=0$, the result will be
\begin{eqnarray}
\label{extresw}
ds^2 &=& dudv +(dx^2)^2 +(dx^3)^2 +\cdots +(dx^{d+2})^2
            +\delta_{ab}dy^ady^b \nonumber \\
		&+&\left [\frac{16\pi p}{(d-1)A_d\ \rho ^{d-1}}
	\right]\delta (u)du^2.
\end{eqnarray}
Note that the ultrarelativistic limit for extremal black $p$-branes
($r_+=r_-$) is also the metric (\ref{extresw}). Therefore, 
the metric (\ref{extresw})  is a gravitational shock wave solution 
generated by a fundamental membrane (string) travels at the speed of 
light. Here the  difference should be noticed between the
solutions (\ref{extresw}) and (\ref{has}). The former is generated by
a membrane (string) and the latter by a point-like particle. For the
solution (\ref{extresw}), the internal space ($\delta_{ab}dy^ady^b$)
is regular and does not relate to  the delta-type singularity.

\subsection{Boosting in the membrane directions}

Now we boost the black $p$-brane along  a direction, say $y^1$, 
of the membrane directions $y$. That is,  we do  the following 
Lorentz transformation 
\begin{equation}
t=\gamma (t'-v y^1), \ \ \ y^1=\gamma (y'^1-v t'),
\end{equation}
and other coordinates keep unchanged. The solution (\ref{bp}) becomes
\begin{eqnarray}
\label{bpp}
ds^2 &=& \gamma^2 [D(r)-A(r)](dt'-v{dy'}^1)^2 +B(r)dr^2 +
   r^2 C(r)d\Omega_{d+1}^2 \nonumber \\
	&+& D(r)[-dt'^2 +({dy'}^1)^2 +\delta_{a'b'}dy^{a'}
	dy^{b'}],
\end{eqnarray}
where $a', b'=2, 3,\cdots,\tilde{d}-1$. Expanding the function
$D(r)-A(r)$, up to the order of  $r_+^{2d}$, $r_-^{2d}$, and $r_+^dr_-^d$,
yields
\begin{equation}
\label{funct}
D(r)-A(r)=(r_+^d-r_-^d)\frac{1}{r^d}+(\gamma_y-1)(r_-^{2d}-r_+^dr_-^d)
\frac{1}{r^{2d}}.
\end{equation}
From (\ref{bp}) it is easy to find  that the  extremal black $p$-brane 
($r_+=r_-)$ is
invariant under the boost along the membrane directions. This can be
understood in terms of the fact that the extremal black $p$-branes (black
string for $p=1$) is the precisely the solutions of the fundamental 
membranes (strings) \cite{horo}.
 For nonextremal black $p$-branes
($r_+\neq r_-$), in order to extract results of physical meaning from 
(\ref{bpp}), we assume that the mass and charge go to zero in the
following way: 
\begin{equation}
\label{mrelation2}
M=\gamma ^{-2}P, \ \ Q=\gamma ^{-2}P_e,
\end{equation}
where  $P$ and $P_e$ are two  constants. Thus the resulting solution
is
\begin{equation}
\label{solution2}
ds^2=dudv +dr^2 +r^2d\Omega_{d+1}^2 +\delta_{a'b'}dy^{a'}dy^{b'}+
\frac{h}{r^d}du^2,
\end{equation}
where $u=y^1-t$, $v=y^1+t$, and the constant $h$ is
$$ h=\frac{8\pi (d+2)P}{(d+1)A_{d+1}}-d\sqrt{\left(\frac{8\pi P}
{(d+1)A_{d+1}}\right)^2
  +\frac{P^2_e}{d^2(d+1)}}.$$
The solution (\ref{solution2}) is a plane-fronted gravitational wave
propagating along the direction $v$. When $Q^2=0$, the constant $h$ 
becomes 
$$h=\frac{16\pi P}{(d+1)A_{d+1}}. $$
In this case, the solution
(\ref{solution2}) is the ultrarelavisitic boost limit of the neutral 
black membrane along the membrane directions. For the neutral 
black string case, the solution (\ref{solution2}) reduces to the 
result given in \cite{horne}.

\section{Boosting the multi-centered black strings and $p$-branes}

In this section we will discuss boosting  the multi-centered black
strings and $p$-branes. Let us consider the most generic multi-centered 
black $p$-brane solution in the following action \cite{khu}
\begin{equation}
\label{action5}
S=\frac{1}{16\pi}\int d^{D+p+q}x \sqrt{-d}\left [R-\frac{1}{2}
 \lambda (\nabla \phi)^2 -\frac{1}{2(p+2)!}e^{a_2\lambda
 \phi}F^2_{p+2}\right ],
\end{equation} 
where $F_{p+2}$ is the $(p+2)$-form tensor field, 
$\lambda =2/(D+p+q-2)$, and
$$ a_2^2=\frac{D+p+q-2}{D-2}a^2 -\frac{(D-2)pq +(D-3)^2p
+q}{D-2}.$$
The multi-centered $(p,q)$-brane solution is \cite{khu}
\begin{equation}
\label{mbp}
ds^2=U^{-2}(r)(-dt^2 +\delta _{\alpha\beta }dy^{\alpha }dy^{\beta})
    +U^{2(p+1)/(d+q)}(r)
    (\delta _{ab}dx^adx^b +\delta _{\mu\nu}dz^{\mu}dz^{\nu}),
\end{equation}
with
\begin{eqnarray}
&& U(r)=H(r)^{(d+q)/[(p+1)(d+q)+a_2^2]},  \nonumber \\
&& H(r)=1+\frac{1}{d}\sum^{n}_{i=1}\frac{\mu_i}{|r-r_i|^d}.
\end{eqnarray}
Here $\alpha, \beta=1,2,\cdots,p$, $\mu.\nu=1,2,\cdots,q$,  
$a,b=1,2,\cdots,d+2$ $(d=D-3)$, and $r_i=(x_i^1,x_i^2,\cdots,
x_i^{d+2})$ is the location of the $i$th membrane. 
The vector potential and dilaton field are
\begin{eqnarray}
&& A=\pm \frac{\sqrt{2a_1}}{H(r)}dtdy^1\cdots dy^p, \\
&& e^{-\phi /a_2}=H(r)^{a_1},
\end{eqnarray}
where $a_1=(d+p+q+1)/[(p+1)(d+q)+a_2^2]$. The mass density of the
$i$th membrane is
\begin{equation}
m_i=\frac{A_{d+1}a_1}{8\pi}\mu_i.
\end{equation}
The multi-centered black ($p,q$)-brane solution (\ref{mbp}) is obviously 
boost invariant in the  $y^{\alpha}$ directions.  We boost therefore the
solution in the $x^1$ and $z^1$ directions, respectively.

(i) {\it In the transverse directions}. Performing the Lorentz
transformation in the $x^1$ direction, we have
\begin{eqnarray}
ds^2 &=&\gamma^{2}[U^{2(p+1)/(d+q)}(\bar{r}')-U^{-2}
       (\bar{r}')](dt'-v dx'^1)^2
     +U^{-2}(\bar{r}')\delta _{\alpha\beta}dy^{\alpha}dy^{\beta} 
	  \nonumber \\
	 & +& U^{2(p+1)/(d+q)}(\bar{r}')[-dt'^2 +(dx'^1)^2
	  +\delta_{a'b'}dx^{a'}dx^{b'} +\delta_{\mu\nu}dz^{\mu}dz^{\nu}],
\end{eqnarray}	  
where $a',b'=2,3,\cdots,d+2$. Taking the limit $v\rightarrow 1$ and
using the relation (\ref{mmrelation}), 
we obtain
\begin{equation}
\label{hmsw}
ds^2=dudv +\delta_{a'b'}dx^{a'}dx'^{b'}+\delta_{\alpha\beta}
     dy^{\alpha}dy^{\beta}+\sum^{n}_{i=1}\frac{d\ p_i \delta(u-u_i)}
	  {(d-1)A_d \rho^{d-1}_i}du^2,
\end{equation}
where $u=x^1-t$, $v=x^1+t$, and 
the definition of $\rho^2_i$ is the same as the one in (\ref{rho_i}).
Obviously, this solution is the generalization of the one for a  single
membrane (\ref{sbp}) and describes the gravitational shock waves generated
by $n$ membranes, which move parallel to each other at the speed of
light. The solution (\ref{hmsw}) is invalid when $d=1$,
i.e., $D=4$. As $d=1$, it should be replaced by
\begin{equation}
ds^2=dudv +\delta_{a'b'}dx^{a'}dx'^{b'}+\delta_{\alpha\beta}
     dy^{\alpha}dy^{\beta}-\sum^{n}_{i=1} 4p_i \ln \rho_i^2 
	  \delta(u-u_i) du^2.
\end{equation}

(ii) {\it In the membrane directions}. Let us boost the solution in the 
    $z^1$ direction. Performing the Lorentz transformation in the $z^1$
	 direction, we have
\begin{eqnarray}
ds^2 &=&\gamma^{2}[U^{2(p+1)/(d+q)}(r)-U^{-2}(r)](dt'-v dz'^1)^2
     +U^{-2}(r)\delta _{\alpha\beta}dy^{\alpha}dy^{\beta} 
	  \nonumber \\
	 & +& U^{2(p+1)/(D+q-3)}(r)[-dt'^2 +(dz'^1)^2
	  +\delta_{ab}dx^{a}dx^{b} +\delta_{\mu'\nu'}dz^{\mu'}dz^{\nu'}].
\end{eqnarray}
We set that the mass density goes to zero as $m_i=\gamma^{-2}P_i$. Thus 
we can arrive at
\begin{equation}
\label{msp}
ds^2=dudv +\delta_{ab}dx^adx^b +\delta_{\alpha\beta}dy^{\alpha}
    dy^{\beta} +\delta_{\mu'\nu'}dz^{\mu'}dz^{\nu'}+\sum^{n}_{i=1}
	 \frac{16\pi}{dA_{d+1}}\frac{P_i}{|r-r_i|^d}du^2,
\end{equation}
where $\mu'$, $\nu'=2,3,\cdots,q$, $u=z^1-t$, and $v=z^1+t$. 
This gravitational wave solution is the
extension of a single  membrane solution (\ref{solution2}) and describes
 $n$ gravitational waves propagating  along the $v$ direction.

\section{Conclusion and discussion}

By boosting the four dimensional and higher dimensional 
dilaton black hole, black $p$-brane (black string for $p=1$),
multi-centered dilaton black hole 
and multi-centered black $p$-brane solutions, we have
investigated in some detail the ultrarelavisitic limit of these
solutions. For the
single dilaton black hole solution and black $p$-brane solution for
which the boost is done along the transverse directions, 
the resulting solutions are  the gravitational shock wave solutions
generated by a single particle and membrane (string) moving at the speed 
of light, respectively. When the charge disappears, the contribution 
of the dilaton field vanishes as well. This point can be seen from the
energy-momentum tensor of the dilaton field. For the multi-centered
dilaton black hole solution and multi-centered black $p$-brane solution 
for which the boost is done  along the transverse directions, the
resulting spacetimes are also shock wave solutions, but produced  by
multiple particles and membranes (strings). These particles or membranes
(strings) move parallel to each other. 
When the boost is made along the membrane directions for the signal 
$p$-brane and multi-centered $p$-brane solutions, the ultrarelativistic 
limits are  the general plane-fronted wave solutions propagating along the
membrane directions. Therefore, the ultrarelativistic limits for black 
$p$-brane solutions are different due to the different boost directions.
The effect of the dilaton field on the ultrarelativistic limit has been
considered. Some peculiar cases appear.  For example, in the $a^2=3/4$ four
dimensional  and $\beta ^2=d/(d+1)$ higher dimensional 
dilaton black holes the contribution of the dilaton field cancels just the
one of the tensor fields. And for $a=1$ four dimensional and
$\beta^2=2d/(d+1)$ higher dimensional dilaton black holes, the rescaling 
relation (\ref{mrelation}) ceases to be valid and one has to use the
relation (\ref{qrelation}) in order to get a distributionally well
defined limit.

It should be pointed out that, to get well defined ultrarelativistic 
limits of dilaton black holes, the mass and charge must go to zero in an 
appropriate way. In this paper we have used the relations
(\ref{mrelation}), (\ref{qrelation}), (\ref{mmrelation}) and
(\ref{mrelation2}), respectively. As for this point, it is worth adding
some remarks here. The rescaling of the mass in (\ref{mrelation}) is
well motivated and saves the energy of the particle from diverging due
to its finite rest mass by keeping the total energy $p$ fixed and
letting $M$ approach zero in the ultrarelativistic limit
\cite{as,stein}. The rescaling of charge in (\ref{mrelation}) is an
unique manner to get a distributionally well defined result and a finite
correction due to the charge. This rescaling of charge is valid for the 
Reissner-Nordstr\"om black hole \cite{lousto1} and $a\ne 1$ four
dimensional dilaton black hole (\ref{met1}) and $\beta ^2 \ne 2d/(d+1)$
higher dimensional dilaton black hole (\ref{met3}). 
For $a=1$  four dimensional and $\beta ^2=2d/(d+1)$ higher
dimensional dilaton black holes, we have to rescale the charge  in the
manner (\ref{qrelation}). In this case, the contribution of the
electromagnetic field vanishes, only the dilaton field makes its 
contribution through the charge. This can be seen from the
energy-momentum tensor of the dilaton, whose first order correction is
proportional to $r^{2d}_-$. However, the resulting fields in  the
rescaling relations (\ref{qrelation}) and (\ref{mrelation}) have  physically
highly unsatisfactory property of a vanishing field but a nonzero,
delta-type energy density, although it is mathematically perfectly
defined \cite{stein}. A physically  intuitive rescaling of charge is 
eq. ~(\ref{mqrelation}). That is the charge goes to zero in the same 
way as that of the mass. In this case, all contributions coming from
the tensor field and dilaton field vanish as the ultrarelativistic limit 
is approached, and resulting geometries are  gravitational shock wave
solutions generated by neutral particles or membranes (strings) 
moving at the speed
of light. In fact, the rescaling relation (\ref{mmrelation}) in the
multi-centered solutions implies that the charge and mass have same
manners going to zero. Therefore the resulting solution is independent
of the charge and dilaton. This rescaling relation may be the basis of
the argument by 't Hooft that the electric charge shifts only the pole
points in the S-matrix of the scattering process \cite{hooft1}. However, 
it should be noticed that, when the boost is along the membrane (string)
direction, the charge has the correction to the resulting solution 
(\ref{solution2}), although the charge has the same rescaling relation as 
the mass (\ref{mrelation2}). At the same time, it should also be
stressed that the boost is just a method to produce a new exact
solution of Einstein's field equations. It changes the type of
solutions, from the type D of the original solutions to the type N of
the new solutions in this paper. From the mathematical point of view,
therefore, the different rescaling relations  of mass and charge are 
admissible.  But the physical relevances need  to be studied  further.

Note that for topological defects as well as point-like sources, the
ultrarelavisitic limit and the weak-field limit (large-distance
behavior) of the original geometries gives the same scattering 
matrices \cite{vega,lousto1,lousto3}. One may expect that it also 
holds for multi-centered black hole or $p$-brane solutions. Compared
with  computing  the scattering matrices in the these solutions, it is 
easier to
calculate the same quantity in the ultrarelavisitic limit of these
solutions. For example, let us consider the scalar field propagating in
the spacetime (\ref{mas}). According to \cite{vega}, for an incoming
plane wave with ingoing momentum (${\bf k}_{\bot},\omega)$,
\begin{equation}
\Psi_<(u,v,{\bf x}_{\bot})=\exp[-i\omega v/4]\cdot 
         \exp [i{\bf k}_{\bot}\cdot {\bf x}_{\bot}-iu ({\bf k}^2_{\bot}
			+m^2)/\omega],
\end{equation}
where ${\bf x}_{\bot}=(y,z)$ and $m$ is the mass of the scalar field,
the outgoing solution with momentum $({\bf p}_{\bot},\omega)$ is given
by
\begin{equation}
\Psi _>(u,v,{\bf x}_{\bot})=\int d^2{\bf p}_{\bot}\exp [i{\bf p}_{\bot}
      \cdot {\bf x}_{\bot}-iu({\bf p}_{\bot}^2+m^2)/\omega]
		S({\bf k}_{\bot},{\bf p}_{\bot},\omega).
\end{equation}
The scattering matrix $S$ is
\begin{equation}
S({\bf k}_{\bot},{\bf p}_{\bot},\omega)=\int \frac{d^2{\bf
x}_{\bot}}{(2\pi)^2} \exp[i({\bf k}_{\bot}-{\bf p}_{\bot})\cdot 
{\bf x}_{\bot} +(i\omega/4)\sum^{n}_{i=1}f(\rho_i)],
\end{equation}
where $f(\rho_i)=4p_i\ln \rho^2_i$. The scattering matrix describes the
scattering process of multiple particles at the energy of the Planckian 
scale, just as discussed in \cite{hooft1,vega,ver}. At the same time, it
also describes the scattering of scalar field off multi-centered black
hole solution under the weak-field approximation. For the scattering 
of string-string, one may study the scattering
of string in the gravitational wave spacetime generated by string moving 
at the speed of light. It would be interesting to further study the
change of poles in these  S matrices.
As an application of shock wave solutions
generated by multiple particles and membranes (strings), one
may use these geometries  
to investigate the scattering processes of
multiple particles and membranes (strings) at the energies of 
Planckian scale. Also it should be of interest to study the geodesics
and geodesic deviation in the shock wave spacetimes generated by
multiple particles and membranes (strings).

As an interesting extension, it is of certain significance to consider 
the gravitational shock waves  generated by multiple particles
and  membranes (strings) in  curved spacetimes.

\section*{Acknowledgments}
R.G. Cai would like to thank C. O. Lousto for helpful email
correspondence, J. H. Cho for fruitful discussions. This work
was supported by the Center for Theoretical Physics of Seoul National 
University.

\end{document}